\documentclass[reqno]{amsart}
\usepackage{graphicx,amsmath, amsaddr}
\usepackage{float}
\usepackage[utf8]{inputenc}
\usepackage{color}
\usepackage{cite}
\usepackage{fancyhdr}
\usepackage[margin=2cm]{geometry}
\usepackage{appendix}
\usepackage{multirow}
\usepackage[justification=centering]{caption}
\floatstyle{plaintop}
\restylefloat{table}
\usepackage{multirow}
\usepackage[table,xcdraw]{xcolor}
\usepackage{amssymb}
\usepackage{siunitx}
\usepackage{chemformula}
\usepackage{booktabs}
\usepackage{physics}

\fancypagestyle{plain}{
	\fancyhf{} 
	\fancyfoot[C]{\thepage} 

}

\title{Thermal and mechanical properties of AlSi7Mg matrix syntactic foams reinforced by Al$_2$O$_3$ or SiC particles in matrix}

\author{A. Feh\'{e}r$^{1,5}$, J. E. Maróti$^{2,3}$, D. M. Takács$^{1,5}$, I. N. Orbulov$^{2,3}$, R. Kovács$^{1,4,5}$}

\address{
	$^1$Department of Energy Engineering, Faculty of Mechanical Engineering, Budapest University of Technology and Economics, Műegyetem rkp. 3., H-1111 Budapest, Hungary \\
	$^2$Department of Materials and Science Engineering, Faculty of Mechanical Engineering, Budapest University of Technology and Economics, M\H{u}egyetem rkp. 3., H-1111 Budapest, Hungary \\
	$^3$ MTA–BME Lendület "Momentum" High-performance Composite Metal Foams Research Group \\
	$^4$Department of Theoretical Physics, Wigner Research Centre for Physics, Institute for Particle and Nuclear Physics, Budapest, Hungary \\
	$^5$Montavid Thermodynamic Research Group, Budapest, Hungary
}
\date{\today}

\begin{document}

\maketitle

\begin{abstract}

Materials with complex inner structure can be challenging to characterize in an effective way. First, it is difficult to determine the representative volume for such a heterogeneous material. Second, the effective material parameters significantly depend on the structure, and thus on the interface properties. In the present paper, we focus on composite metal foams, and we attempt to determine the effective thermal parameters. We use the heat pulse experiment to study its transient thermal response. We observed deviation from Fourier's law, and thus we propose an evaluation procedure for such experimental data using the Guyer--Krumhansl heat equation. Furthermore, we studied the effective parameters, the specific heat, mass density, thermal conductivity and thermal diffusivity, and we propose a method for deducing a more reliable thermal diffusivity and thermal conductivity based on the transient temperature history.

\end{abstract}

\maketitle

\section{Introduction}
Metallic foams are a versatile group of materials, widely applied in e.g.~the
automotive and transportation industry~\cite{banhart_aluminium_2005,song_experimental_2020,srinath_characteristics_2010,linul_axial_2021,claar_ultra-lightweight_2000,simancik_metallic_2001},
in noise and sound management~\cite{lefebvre_porous_2008,jones_assessment_2009},
and in the aerospace industry~\cite{liu_prediction_2012,klavzar_protective_2015,williamsen_video_2001,ryan_hypervelocity_2010}.
Their commonly exploited, unique mechanical properties include high specific strength and high mechanical energy absorption capability~\cite{rajak_insight_2020,gupta_metal_2014}. One of the key factors influencing the performance of metallic foams is their internal structure, based on which they are usually grouped into three main categories: open cell, closed cell and metal matrix syntactic foams (MMSFs). MMSFs are composite materials where inside the metal matrix a certain hollow or porous phase filler is introduced, such as ceramic hollow spheres (CHSs), glass microspheres, clay particles or fly ash cenospheres, among many others~\cite{gupta_metal_2014,cochran_ceramic_1998,wright_processing_2017,szlancsik_mechanical_2020}. Additionally, the matrix material can also be reinforced by adding various particles~\cite{stevenson_particle_2012}. This approach has originally been introduced for traditional metal foams, and has only been applied recently to MMSFs~\cite{maroti_characteristic_2023}.

Furthermore, beyond their advantageous mechanical properties, metallic foams can also be used in thermal engineering applications such as heat exchangers, cooling and thermal storage systems~\cite{boomsma_metal_2003,ejlali_application_2009,jannesari_experimental_2017,chen_thermal_2021,talebizadehsardari_consecutive_2021}. This is due to the fact that, e.g.,~Al and Cu metal foams have a high thermal conductivity, while a porous structure also allows for fluids or gases to pass through them. In thermal storage systems or heat exchangers where phase change materials (PCMs) are used, the unique material structure of the metallic foams allows for significantly improved performance, as their good thermal conduction and high internal surface area complements the poor thermal conduction of the high latent heat capacity PCM inside the pores~\cite{chen_thermal_2021,shu_effect_2023,tian_numerical_2011,nemati_pore-scale_2023}.

For use of metal foams in thermal engineering applications, a reliable and accurate thermal model of them is needed. However, their characteristic heterogeneous structure poses a difficulty here, as the traditional Fourier heat conduction model is often insufficient to describe transient heat conduction in heterogenous media such as metal-polymer composites and rocks due to appearance of the so-called over-diffusion \cite{both_deviation_2016,van_guyer-krumhansltype_2017,fulop_emergence_2018,feher_evaluation_2021,lunev_digital_2022,nemati_pore-scale_2023}. According to our experience, such deviation can only be apparent in a transient situation. Steady-state temperature distributions, therefore the static effective thermal conductivity measurements are not expected to be distorted by over-diffusion. However, the effective thermal conductivity determined in a static measurement is not necessarily applicable in a transient case for such materials as their dynamic behavior is strongly influenced by the interaction of multiple heat transfer channels.

One approach to solve this is the use of microscopic or pore-scale simulations~\cite{tian_numerical_2011,lunev_digital_2022,nemati_pore-scale_2023}, which, while providing accurate results, come with several technical difficulties. For example,~\cite{lunev_digital_2022} shows that in an open-cell metal foam, only a very high-resolution geometric model, such as one produced by an X-ray computer tomography scan, can accurately describe thermal behaviour if the Fourier model is used. Consequently, the computational demand of doing calculations using microscopic models is very high. Clearly, while using microscopic simulations allow for better understanding of the underlying physical processes, their use in everyday engineering practice is less favourable.

On the other hand, continuum models beyond Fourier's law can provide a faster and more concise description of the thermal material behaviour while using effective bulk parameters. The fundamental difference between thermally homogeneous and heterogeneous materials is the number of separate channels of heat conduction: while in the former only a single microscopic channel of heat conduction exists, in the latter the heterogeneity can create an additional path for heat conduction with different properties. Of the several available models employed for the thermal characterization of heterogeneous media, the Guyer--Krumhansl (GK) model~\cite{guyer_thermal_1966} has shown the most success~\cite{both_deviation_2016,feher_analytical_2022}. This success is partly due to the GK model being able to model two separate thermal timescales, corresponding to the separate channels of heat conduction in heterogeneous media. The GK model also shows thermodynamic consistency~\cite{van_guyer-krumhansltype_2017,fulop_emergence_2018}, unlike other models, e.g., the delay-type dual phase lag equation \cite{fabrizio_stability_2014}. As a continuation of this work, the thermal conductive behaviour of MMSFs is investigated in this paper in both the Fourier and non-Fourier paradigm, through the GK model. To the best of our knowledge, the applicability of the GK model on MMSFs for an accurate description of their thermal behaviour has not been published before in the literature.

This study mainly focuses on the thermal characterization of AlSi7Mg matrix syntactic foams with different reinforcement particles in the matrix, while the production and mechanical characterisation of the specimens are also included briefly for the sake of a more complete understanding. Section 2 details the various materials used in the production of the reinforced metal foam samples, including a brief explanation of the production process. Section 3 deals with the mechanical characterization of the samples, including the determined compressive strength, plateau strength, energy absorption, structural stiffness and energy absorption efficiency values. For the compressive strength and the structural stiffness, two linear relationships are provided as a function of the reinforcing particle size in the matrix. Section 4 details the thermal characterization of the metallic foam samples by heat pulse experiments. First, a thermal model for the performed experiments is introduced (Section 4.1), followed by an in-depth explanation of the experiment evaluations (Section 4.2). After evaluating the measurement data, the resulting thermal parameters are given (Section 4.3), including the Fourier thermal diffusivity, Guyer--Krumhansl thermal diffusivity, heat flux relaxation time and spatial scale parameter. Section 4.5 gives a brief comparison of measurement results and effective thermal conductivity values calculated using estimation formulae from literature. Finally, the results are discussed in Section 5.
\section{Materials and production}

\subsection{Materials of the reinforced metal foam samples}

In this study, A356 (AlSi7Mg) has been used as matrix material. Ceramic hollow spheres (CHSs) were used as the porosity-ensuring filler phase, as shown in Fig.~1.

The CHSs' diameter was $\varnothing 2.27 \pm \SI{0.13}{mm}$, based on microscopic image measurements of numerous (100 randomly selected) particles, and their calculated bulk density was $0.94 \pm \SI{0.05}{g.cm^{-3}}$. The material of the CHSs was pure \ch{Al2O3}.

\begin{figure}[H]
	\centering
	\includegraphics[width=15 cm]{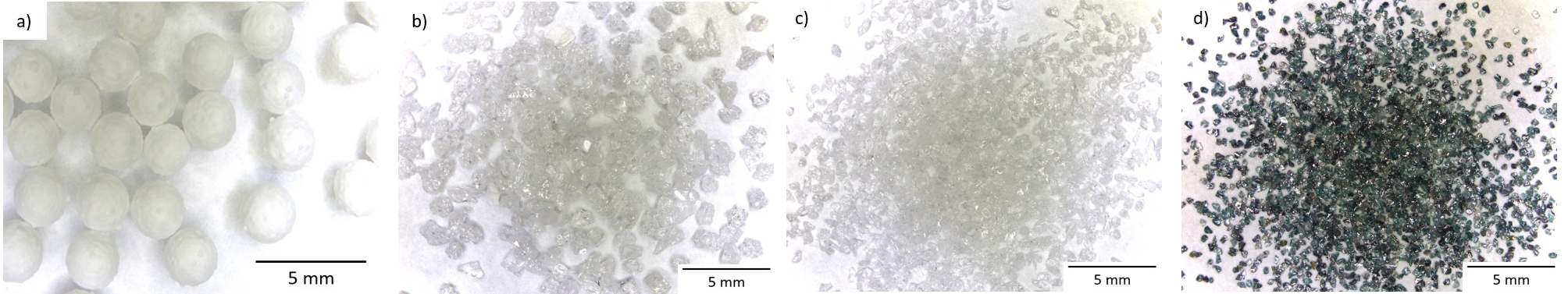}
    \caption{Investigated filler and reinforcing materials, a) CHSs, b) \ch{Al2O3} (with \SI{1.2}{mm} nominal size), c) \ch{Al2O3} (with \SI{0.6}{mm} nominal size),  d) \ch{SiC}}
	\label{fig:filler}
\end{figure}

Three different particles were used as reinforcing materials: two different sizes of alumina, pure \ch{Al2O3} (with 1.2 mm and 0.6 mm nominal size), and silicon-carbide, pure SiC (0.4 mm nominal size). For each case, 20 vol\% of reinforcing particles were used, which is the quantity of reinforcing particles relative to the volume of the matrix material (corresponds to 7 vol\% of the whole MMSF). The reinforcing particles are also shown in Fig.~\ref{fig:filler}.

A summary of types, sizes, and combinations of filler materials and reinforcing particles is listed in Table 1. In the designation, the first part describes the applied reinforcement (A stands for Al$_2$O$_3$ particles and S for SiC particles, respectively) followed by the nominal size of the particles, and the last part refers to the filler material, for example A-1.2-CHS.
\begin{table}[H]
\centering
\caption{Designations of the specimens and summary of filler materials and reinforcing particles}
\begin{tabular}{c c c c c c}
\toprule
\multirow{2}{*}{Sample ID} & \multirow{2}{*}{Designation} & \multicolumn{2}{c}{Filler material} & \multicolumn{2}{c}{Reinforcing particles} \\
\cmidrule{3-6}
 & & Type & Diameter -- $D$ [mm] & Composition & Measure size -- $d$ (mm) \\
\midrule
\midrule 
22; 23; 24 & S-0.4-CHS & CHS & $\varnothing 2.27 \pm 0.13$ & SiC & $0.41 \pm 0.11$ \\
32; 33; 34 & A-0.6-CHS & CHS & $\varnothing 2.27 \pm 0.13$ & \ch{Al2O3} & $0.64 \pm 0.21$ \\
42; 43; 44 & A-1.2-CHS & CHS & $\varnothing 2.27 \pm 0.13$ & \ch{Al2O3} & $1.22 \pm 0.19$ \\
\bottomrule
\end{tabular}
\label{table_1}
\end{table} 

\subsection{Production of the reinforced AlSi7Mg matrix syntactic foams}

The specimens were produced by low pressure infiltration. The CHSs and the reinforcing particles were mixed until the mixture was homogenous in the production process. After the mixture was placed into the mold, a stainless-steel mesh was inserted atop the mixture, which was wedged against the walls of the mold to prevent the movement of the filler material.
 
After preheating the prepared molds for \SI{30}{min} at \SI{600}{\degreeCelsius}, the aluminum melt at minimum \SI{840}{\degreeCelsius} was poured onto the top of the mixture and then pressed between the particles using argon gas at a pressure of \SI{500}{kPa}. The infiltration time was 5 sec. The air under the melt exited the mold through a condensed tube at the bottom.

Heat treatment was applied to the specimens to reach the T6 treatment state. The first step is heating at \SI{300}{\degreeCelsius/h} rate to \SI{535}{\degreeCelsius}, holding for 4 hours, then quenching in water. The second step was heating at \SI{200}{\degreeCelsius/h} rate to \SI{150}{\degreeCelsius}, holding for 15 hours, and then quenching again in water.

\section{Mechanical characterization and properties}

Prior to the thermal experiments, compressive tests were performed on the samples. The detailed results were published in~\cite{maroti_characteristic_2023}, but a summarized version of the compressive tests' results is published here for better understanding.

Compression tests were performed with an MTS810 universal hydraulic testing machine between two flat steel platens. The device was equipped with a 250 kN load cell. A 0.3 mm thick Kolofol Teflon foil was used for lubrication, placed on the contact surfaces between the specimens and the platens. The tests were carried out on 12 specimens. Each specimen was compressed with a 4 mm/min cross-head speed to at least 50\% engineering strain value for comparability, based on ISO13314:2011~\cite{isotc_164sc_2_ductility_testing_technical_committee_mechanical_2011}.

The following characteristic mechanical properties were evaluated from the results: compressive strength ($\sigma_c$ - compressive stress at the first peak after the elastic deformation), plateau strength ($\sigma_{pl}$ – the average stress in the range of 10\%–40\% deformation), energy absorption ($W_{50\%}$ – area under the engineering stress-engineering strain curve up to 50 \% strain), structural stiffness ($k$ – the slope of the linear part of the engineering stress-engineering strain curve) and energy absorption efficiency ($W_{\eta}$ – absorbed energy divided by the energy absorbed by an ideal metal foam).

After the examination of the results, the following conclusions and findings can be drawn.

The shape of the engineering stress–engineering strain curves was mainly influenced by the choice of the filler. CHS filled MMSFs showed a high-stress peak after the initial linear elastic part of the curves and provided high plateau stress level.

The compressive strength and the structural stiffness of the MMSFs were increased by the reinforcing particle in the matrix. A linear relationship has been observed between the $Al_2O_3$ particle size $d$ and the structural stiffness $k$, in the form of~\eqref{eq_slope}. A similar relationship can be fitted to the compressive strength $\sigma_c$ as a function of $d$, which is expressed as~\eqref{eq_sigmapl}.

\begin{align} 
    k &= \SI{51.6}{MPa} - \SI{8.2}{MPa/mm} \cdot d; \quad R^2 = 0.907,\label{eq_slope}\\
    \sigma_c &= \SI{102.6}{MPa} + \SI{23.7}{MPa/mm} \cdot d; \quad R^2 = 0.981\label{eq_sigmapl}.
\end{align}

Additionally, the presence of the reinforcement decreased the plateau strength and the absorbed mechanical energy (there is a strong connection between the energy absorption and the plateau strength) in the case of Al2O3 particles, and the decreasing effect caused by the stress concentrating particles could be only balanced by the stronger SiC particles.

\section{Thermal characterization and properties}
For thermal characterization, heat pulse experiments were performed. As shown in prior works, heterogeneous materials can exhibit non-Fourier heat conduction at room temperature, for which the more sophisticated, thermodynamically consistent Guyer--Krumhansl (GK) model can usually provide an adequate description~\cite{both_deviation_2016,van_guyer-krumhansltype_2017,fulop_emergence_2018}. Thus, in the following, both the classical Fourier model and the GK model are fitted to the measurements.

For determining the parameters based on the experiments, an iterative procedure is used, involving sensitivity functions, being particular for each model. Therefore the iteration procedure includes the model characteristics encoded in the sensitivity functions, and iterates in all parameters simultaneously. This ensures that both heat conduction models can be efficiently fitted to the measurements in the best way possible. Naturally, the maximal accuracy is limited by the model used: in case of non-Fourier heat conduction, the best possible fit of the Fourier model will still be unable to accurately explain the thermal behaviour. This demands a more complex constitutive relation, which is satisfactorily provided by the GK model.

\subsection{Experimental setup}
\begin{figure}
	\centering
	\includegraphics[width=7 cm]{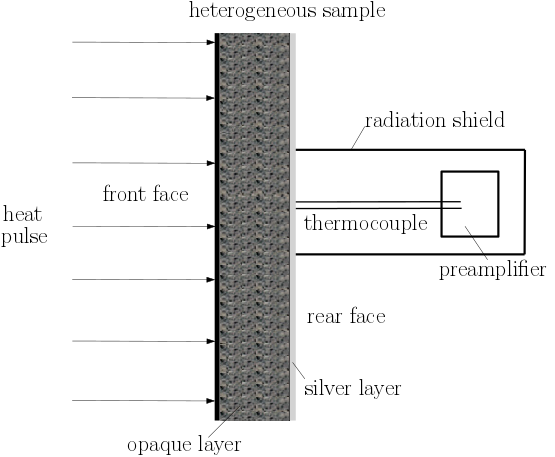}
	\caption{Arrangement of the experimental setup for the heat-pulse measurement.}
	\label{fig:arrangement}
\end{figure}

The heat pulse experiments performed aim to characterise the samples' thermal diffusive behaviour. In the experiments, a flash lamp generates a short heat pulse ($0.01$ s) at the front face of the sample, while the temperature is measured by a pin-thermocouple (K-type) at the rear face, similarly to the experiments detailed in~\cite{both_deviation_2016,van_guyer-krumhansltype_2017,fulop_emergence_2018}. The experimental layout is schematically shown in Fig.~\ref{fig:arrangement}.

For correct measurements, meticulous sample preparation is required. This is complicated since the samples to be measured are required to be very thin, on the order of a few millimetres due to the limited capability of the measurement device. This is usual for any standardised equipment, and also strongly limits the possibilities about experimenting on heterogeneous materials with macroscale inclusions. Producing the same thicknesses in this size range, especially for metal foams, is a challenge. Ideally, in order to obtain more representative thermal characteristics for a given sample, a much larger specimen would have to be tested for which the interaction of parallel heat transfer channels becomes less significant and does not influence the transient behavior. However, the use of such small samples with a thickness between $2$ and $3$ mm, can reveal the more detailed behavior of a structure, and thus deepen our understanding of the effective characterization of complex structures.

\subsection{Models for heat pulse experiments}
In order to evaluate the measurements, a thermal model for the experimental setup is needed.
Since the thickness of the specimens is significantly smaller than their diameter, besides the heat pulse homogeneously excites the entire front face, a one-dimensional model can be used for this experiment, with a time-dependent heat flux boundary condition at one end, and a heat transfer boundary condition at the other. Notably, while one can define heat transfer boundary condition also on the front and lateral faces of the specimen with different heat transfer coefficients, these extra parameters cannot improve the fitting. This is reasonable, since a few seconds after the heat pulse entered the specimen, the essential transients are ended, followed by a slow natural cooling period. During that period, the specimen behaves as a lumped, point-like body without resulting in a notable temperature difference inside. Consequently, one heat transfer boundary condition is enough. Additionally, as the temperature changes during the experiments are low (3--\SI{5}{K}), temperature dependence of the material parameters is not considered. 

Assuming a homogeneous material, the Fourier constitutive relation in one dimension reads as
\begin{gather}
    q(x,t)=-\lambda \pdv{T(x,t)}{x},\label{eq:fourier_constitutive}
\end{gather}
where $T(x,t)$ is the temperature in the one-dimensional domain at time $t$ and position $x$, $q(x,t)$ is the heat flux, and $\lambda$ is the thermal conduction coefficient. As an extension of this relation, the GK constitutive model can be written as
\begin{gather}
    \tau_q \pdv{q(x,t)}{t} + q(x,t) = -\lambda \pdv{T(x,t)}{x} + \kappa^2 \pdv[2]{q(x,t)}{x},\label{eq:gk_constitutive}
\end{gather}
where $\tau_q$ is the heat flux relaxation time and $\kappa$ can be interpreted as a spatial scale parameter. We want to note here that originally, the GK equation is derived in the framework of phonon hydrodynamics, and thus such model would be restricted to low-temperature situations due to the prescribed propagation mechanism. In such a model, this becomes apparent by the pre-determined transport coefficients. For instance, in a phonon hydrodynamic model, neither the thermal conductivity, nor the $\kappa$ spatial scaling parameter can be fitted.
However, in our situation, we consider the continuum background of the GK equation. In other words, there is no any prior assumption on the microscopic propagation mechanism, and all coefficients in Eq.~\eqref{eq:gk_constitutive} must be determined from measurements with a proper fitting procedure.

These constitutive equations are accompanied with the balance of internal energy,
\begin{align}
 \rho c_v \frac{\partial T }{\partial t} = - \frac{\partial q}{\partial x},
\end{align}
in which $c_v$ is the isochoric specific heat assuming a rigid conductor, and we omit any heat sources, therefore we also assumed that the heat pulse completely absorbed on the front boundary. Consequently, the boundary conditions can be expressed as
\begin{align}
    q(x{=}0, t) &= q_0(t), \\
    q(x{=}L, t) &= - h \big( T(x{=}L, t) - T_{\infty} \big),
\end{align}
where $L$ is the thickness of the sample, $h$ is the heat transfer coefficient, $T_{\infty}$ is the constant ambient temperature, and $q_0(t)$ describes the heat flux from the flash lamp.

For constitutive relations~\eqref{eq:fourier_constitutive} and~\eqref{eq:gk_constitutive}, including the balance of internal energy and eliminating either the heat flux or the temperature field from the equations, the $T$- or $q$-representation of the corresponding model can be obtained, respectively (c.f.~\eqref{eq:GK_T_representation}). Afterwards, for the evaluation of the experimental measurements, the equations can be solved for the rear side temperature as a function of time: this is detailed in~\cite{feher_evaluation_2021,feher_analytical_2022}.

For practical reasons, the equations used in evaluating the measurements are brought to a dimensionless form, most notably the quantities below are nondimensionalized in the following way:
\begin{gather}
    \tilde{x} = \frac{x}{L},\quad \tilde{T} = \frac{T - T_{0}}{T_{\infty} - T_{0}}
\end{gather}
with $T_0$ being the initial (homogeneous) temperature of the sample; see~\cite{both_deviation_2016} for more details. In the following, dimensionless quantities are used (with the tilde notation omitted), unless indicated otherwise.

We also want to highlight the basic difference between the Fourier and GK equations by presenting their analytical solutions for such an experimental setting. The usual one-term solution (i.e., the first term of the infinite series) for the rear side temperature for the Fourier model is
\begin{gather}
    T(x{=}1,t) = Y_0 \exp\left( -h t \right) - Y_1 \exp\left( x_F t \right),\quad \textnormal{with} \quad  x_F=-2h -\alpha \pi^2 <0, \label{a1}
\end{gather}
where $x_F$ characterises the Fourier conduction time scale with the dimensionless thermal diffusivity $\alpha$, and that scale is also influenced by the heat transfer on the boundary, characterized by $h$. On the other hand, the GK solution highlights an additional timescale compared to the Fourier solution:
\begin{gather}
    T(x{=}1,t) = Y_0 \exp\left( -h t \right) - Z_1 \exp\left( x_1 t \right) - Z_2 \exp\left( x_2 t \right),\quad \textnormal{with} \quad x_1, x_2 < 0, \label{a2}
\end{gather}
as shown and detailed in~\cite{feher_evaluation_2021,feher_analytical_2022}. For the exponents, $x_2<x_F<x_1$ holds. In the iteration procedure, a more complete analytical solution is used, i.e., we utilized $100$ terms from the infinite series to construct the rear side temperature history.

It is worth emphasizing that even though the MMSF material considered here is obviously non-homogeneous, the small-scale variations in the material cannot be modelled realistically in a practical, engineering use-case. Thus, a bulk description of the material is needed, where the additional complications introduced by the metal foam topology are modelled by additional terms in the constitutive relation, as in the case of the GK model. By introducing a more sophisticated material model, the bulk parameters can be kept homogeneous, one can avoid the use of detailed, highly computationally intensive simulations.

\subsection{Evaluation of heat pulse measurements}
From the heat pulse measurements, a temperature time series is obtained, which can be used to fit the solutions of both the Fourier and the GK models. The most important parameter, appearing in both models, is the thermal diffusivity $\alpha$. In contrast to the Fourier model, the GK model also contains two additional parameters.

For achieving the best fit for a given model, an iterative procedure is used leveraging sensitivity functions derived from the dimensionless model equations, yielding dimensionless parameters. These can be used to recover the dimensioned parameters sought. It is worth noting that the dimensionless heat transfer coefficient of the experimental setup must also be fitted, but for recovering the actual value of $h$ would the value specific heat would also be needed.

\subsubsection{Iterative determination of parameters}
By iterative procedure, we mean that we look for the parameters in the given parameter space that best fit the given model over the entire time series. This is not equivalent to achieving the best fit physically. A good example of this is when we want to fit the Fourier equation to a data set that only after fitting turns out not to be fully explained by the Fourier equation.

In this case, although we may achieve the best available fit in terms of $R^2$, it may not reflect our physical expectations. The difference between the Fourier and GK equations is again due to the time scales. 

The natural expectation is that, as the initial significant transient effects fade, the differences due to the heat conduction mechanisms will dissipate, and if the process is sufficiently slow, we will see no difference in the temperature-time data predicted by the two different models. We call this phenomenon Fourier resonance, which can be observed from the $T$-representation of the GK equation:
\begin{align}
    \Big (\partial_t T + \alpha \partial_{xx} T \Big) + \tau_q \partial_t\Big( \partial_t T + \frac{\kappa^2}{\tau_q} \partial_{xx} T \Big) = 0,\label{eq:GK_T_representation}
\end{align}
in which the Fourier equation and its time derivative are also present, such that when $\alpha = \kappa^2/\tau_q$, we obtain exactly the solution of the Fourier equation.

We see this property experimentally, and the model parameters imply that
\begin{align}
	\alpha_F \approx \frac{1}{2} \left ( \alpha_{\textrm{GK}} + \frac{\kappa^2}{\tau_q}  \right),
\label{eq:atlag}
\end{align}
i.e., the heat conduction coefficient given by the Fourier equation is the average of the parameters of the GK equation, which thus shows the time scale averaging. More importantly, Eq.~\eqref{eq:atlag} can offer a helpful approximation for Fourier's thermal conductivity if the parameters of the GK equation are determined. It means that such $\alpha_F$ can be used to predict the long-term (or sufficiently slow) transient behaviour of the complex metal foam structures. 

For this reason, when a "physically correct" Fourier fit is sought globally for the whole time series on a non-Fourier data series, the cooling phase is given as a strong constraint and the thermal diffusivity will converge to an optimal value in this respect. 

\subsubsection{Sensitivity functions}
The iteration is based on the so-called local sensitivity functions, i.e.
\begin{align}
	S_{p_i,t} = \frac{\partial y}{\partial p_i}, \quad y=y(p_i, t), \quad i=1,\dots,N,
\end{align}
in which $y(p_i, t)$ denotes the time series by model prediction, hence this is the rear side  temperature in our case, which depends on $N$ number of parameters $p_i$. Thus, the sensitivity function $S_{p_i}$ specifies how a given temperature profile changes in time for a given parameter change in the function $y$ to be predicted in the neighborhood of the parameter $p_i$. That procedure is not only valid for temperature-time series, but is a completely general methodology. It is also worth mentioning reduced sensitivity functions, 
\begin{align}
	\hat S_{p_i,t} = p_i \frac{\partial y}{\partial p_i}, \quad y=y(p_i, t), \quad i=1,\dots,N,
\end{align}
which makes the functions $\hat S$ comparable by bringing them to the same dimension by multiplying by the parameter $p_i$. According to the~\cite{le_masson_metti_2015} literature, the rule of thumb is that if the functions $\hat S_{p_i,t}$ are identical or nearly identical, then their parameters are linearly dependent, so that they cannot be fitted simultaneously. 
We want to highlight that these sensitivity functions are valid only locally, which means that the sensitivity $S$ also depends on which value of $p_i$ takes since our models are nonlinear with respect to the fitted parameters, both the exponents and the coefficients in Eqs.~\eqref{a1}-\eqref{a2} depend on the unknown thermal parameters. Thus it is inevitable to determine the corresponding sensitivity functions in each iteration for each parameter set.

\begin{figure}
	\centering
	\includegraphics[width=7.6cm]{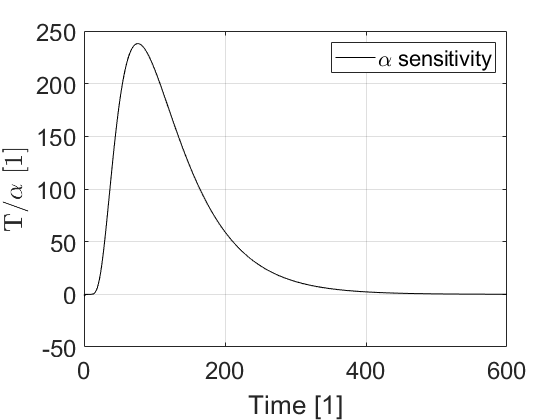}
	\includegraphics[width=7.6cm]{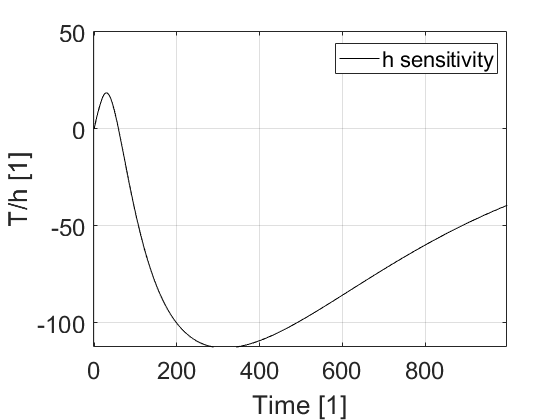}
	\caption{Sensitivity curves for the Fourier equation.}
	\label{fig:F_erzekenyseg_nemred}
\end{figure}

\begin{figure}
	\centering
	\includegraphics[width=7.6cm]{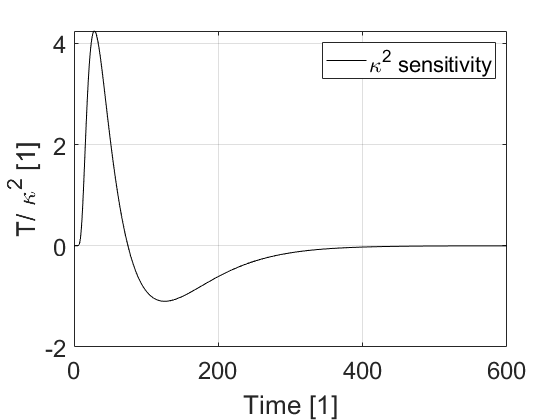}
	\includegraphics[width=7.6cm]{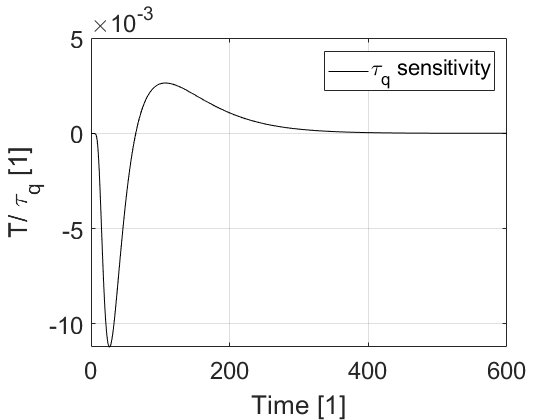}
	\caption{Sensitivity curves for the GK equation.}
	\label{fig:GK_erzekenyseg1_nemred}
\end{figure}

\begin{figure}
	\centering
	\includegraphics[width=7.6cm]{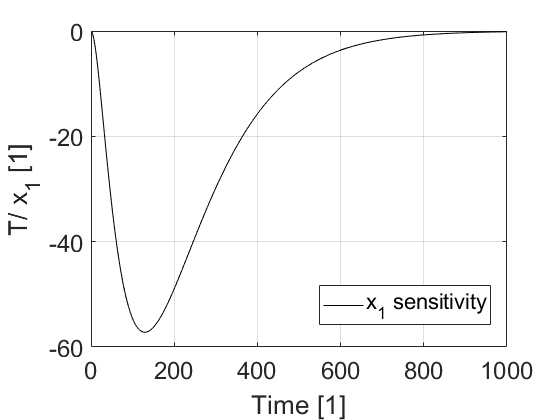}
	\includegraphics[width=7.6cm]{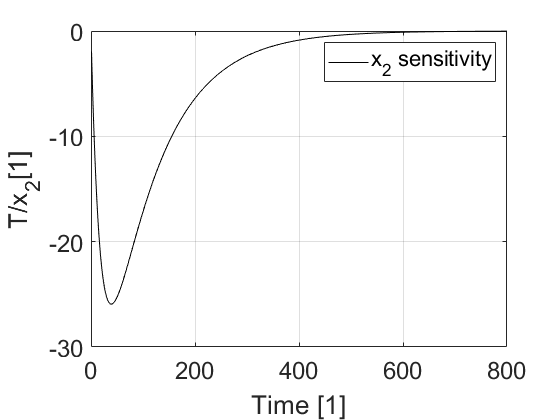}
	\caption{Sensitivity curves for the GK equation.}
	\label{fig:GK_erzekenyseg2_nemred}
\end{figure}

\begin{figure}
	\centering
	\includegraphics[width=7.6cm]{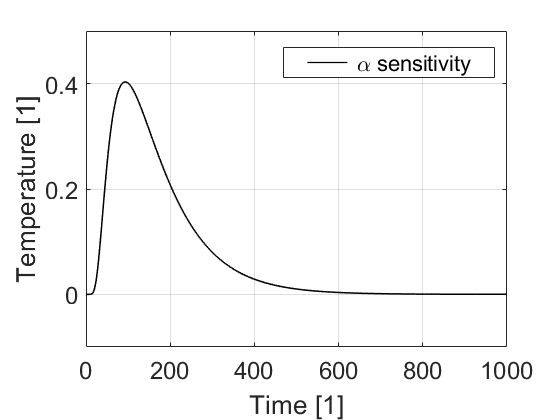}
	\includegraphics[width=7.6cm]{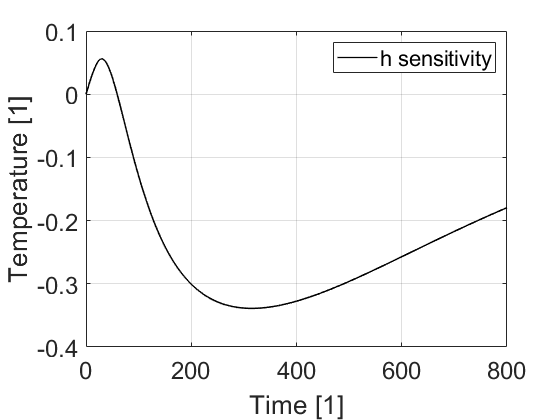}
	\caption{Reduced sensitivity curves for the Fourier equation.}
	\label{fig:F_erzekenyseg}
\end{figure}

\begin{figure}
	\centering
	\includegraphics[width=7.6cm]{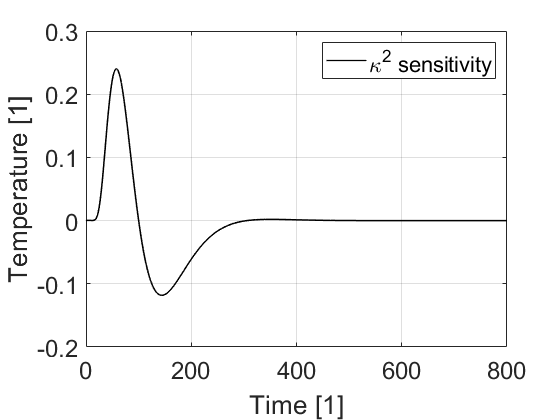}
	\includegraphics[width=7.6cm]{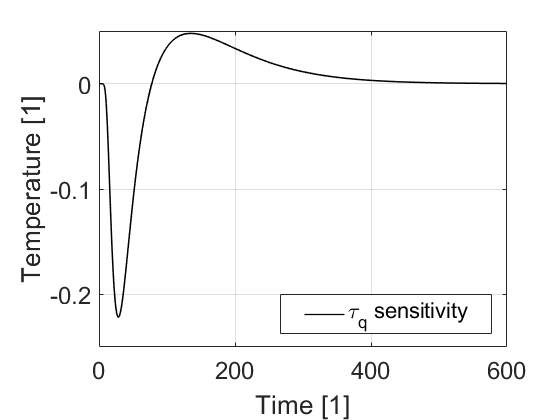}
	\caption{Reduced sensitivity curves for the GK equation.}
	\label{fig:GK_erzekenyseg1}
\end{figure}

\begin{figure}
	\centering
	\includegraphics[width=7.6cm]{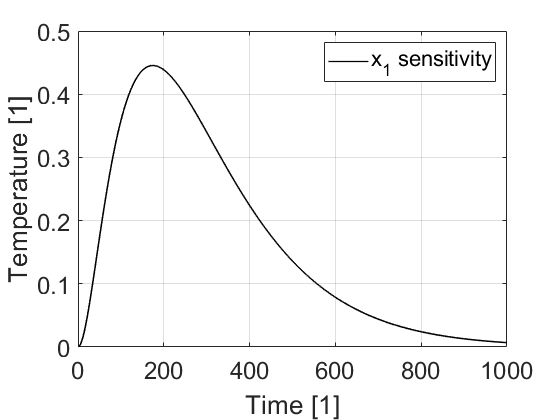}
	\includegraphics[width=7.6cm]{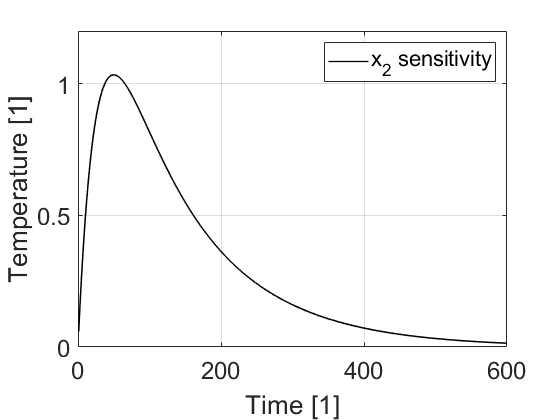}
	\caption{Reduced sensitivity curves for the GK equation.}
	\label{fig:GK_erzekenyseg2}
\end{figure}

\begin{figure}
	\centering
	\includegraphics[width=7.6cm]{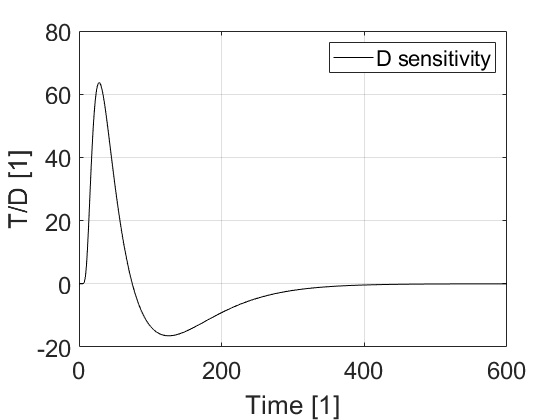}
	\includegraphics[width=7.6cm]{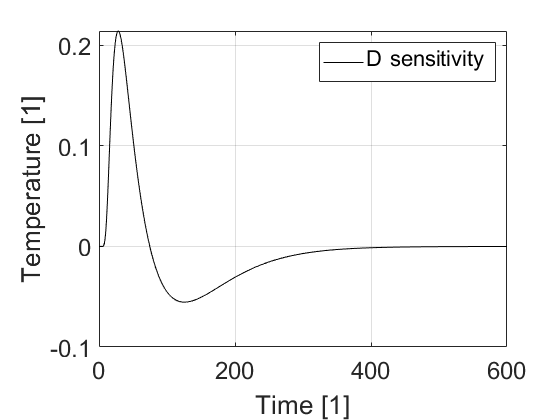}
	\caption{Sensitivity and reduced sensitivity curves for the $D$ parameter for the GK equation.}
	\label{fig:GK_erzekenyseg3}
\end{figure}

In Fourier's case, the unknown parameters are $\alpha$ and $h$. For the GK equation, they are complemented by $\tau_q$ and $\kappa^2$. 
Given that neither the Fourier nor the GK equations are nonlinear in their parameters due to exponential time decay, and thus require an iterative approach, the sensitivity functions $S_{p_i}$ cannot be expressed in a reasonably simple analytical form, and are therefore derived using the following approximation \cite{feher_analytical_2022},
\begin{align}
	S_{p_i,t} = \frac{\partial y}{\partial p_i} \approx \frac{T(p_i+\Delta p_i,t)   -T(p_i,t)}{\Delta p_i}, \quad  \Delta p_i = 0.05 p_i, \label{eq4}
\end{align}
where we treat the partial derivative in a similar way to the forward finite difference approximation. We impose a 5\% difference between the parameters to determine $\Delta p_i$ for a given value of $p_i$. Although the sensitivity functions can dramatically change in the GK's case, they remain characteristic when $\alpha > \kappa^2/\tau_q$ and $\alpha < \kappa^2/\tau_q$. From an experimental point of view, only the $\alpha < \kappa^2/\tau_q$ region is important, this is characteristic for over-diffusion. Figs.~\ref{fig:F_erzekenyseg_nemred}--\ref{fig:GK_erzekenyseg3}.~for the Fourier and GK equations illustrate the corresponding sensitivity and reduced sensitivity functions for each typical parameter. 

\subsubsection{The structure of the iteration}

The sensitivity functions specify the interval of the time series at which a given parameter is most sensitive, so that deviations at that interval are said to be more significant. Thus, iteration based on sensitivity functions searches for the optimal set of parameters based on the properties of the model. This is coupled with a less advantageous property that the sensitivity functions can differ by several orders of magnitude, which also characterises the parameter determinability in some respects. The less sensitive the model to a change in a parameter, the more difficult it is to find its optimal value, and the less that parameter will dominate the iteration procedure used for fitting. 

The iteration is performed for the parameters $h$, $\alpha$, $\tau_q$ and $\kappa^2$, but we find that the sensitivity of $\tau_q$ and $\kappa^2$ is of different by three orders of magnitude (Fig.~\ref{fig:F_erzekenyseg_nemred}-\ref{fig:GK_erzekenyseg2_nemred}). To overcome this difficulty, and to bring the orders of magnitude closer together, we introduce the parameter $D$=$\frac{\kappa^2}{\tau_q}$ instead of $\kappa^2$ in which the iteration is performed (see Fig.~\ref{fig:GK_erzekenyseg3}). Thus, in the evaluation method, the iteration is performed on the parameters $\tau_q$, $h$, $\alpha$ and $D = \frac{\kappa^2}{\tau_q}$.

For a given set of local sensitivity function $S_{p_i}$, we can construct a sensitivity matrix $\mathbf S$, such that $\mathbf S = [S_{p_1} S_{p_2} \dots S_{p_N}]$, i.e., it has as many columns as the number of parameters we wish to iterate over, and as many rows as the length of the time series (number of time steps we apply) over which we are testing the sensitivity. 
Let $P_k$ denote the $k$th iteration of the parameters to be fitted, so that
\begin{align}
	P_{k+1} = P_k + (\mathbf S_k^T \mathbf S_k)^{-1} \mathbf S^T_k (T_{\textrm{measured}} - T (P_k)),
\end{align}
which iteration is non-convergent if any column of the matrix $\mathbf S$ is linearly dependent, i.e., any two parameters are linearly dependent, otherwise the matrix $(\mathbf S_k^T \mathbf S_k)$ becomes singular, thus non-invertable.

The resulting error curve $T_{\textrm{diff}}=T_{\textrm{measured}} - T (P_k)$, in addition to the usual $R^2$, further characterizes the goodness of fit and hence the parameters, as well as the measurement setup itself using a reference sample.

While $T_{\textrm{diff}}$ can be characterized by its $\bar \varepsilon$ mean and $\sigma^2$ variance, we can provide a feedback to the parameters to be fitted via the reduced sensitivities of $\hat S$. If the reduced sensitivity of a given parameter is smaller than the error curve $T_{\textrm{diff}}$ (especially if it is smaller at any point in the entire time series), or its average value, then that parameter is very poorly measured by the instrument, since the errors from the measurement are larger than the local effect of that parameter, and hence the corresponding parameter error is more significant. Hence the proper choice of parameter set can greatly aid the fitting process. 
Naturally, the opposite is also true, if the reduced sensitivity is much larger than the error curve along the entire duration, then the value of the parameter can be well determined even with measurement errors. 

The errors for each parameter can be obtained using the so-called "information matrix" compiled from the variance and the reduced sensitivity functions,
\begin{align}
	\mathbf C = \sigma^2 (\hat{\mathbf S}^T \hat{\mathbf S})^{-1}, \quad \hat{\mathbf S}= [\hat S_{p_1} \hat S_{p_2} \dots \hat S_{p_N}].
\end{align}

This has the additional advantage that the $\hat{\mathbf S}$ matrix is clearly a feature of the model, and larger sensitivity amplitudes can also magnify measurement errors. Together, the root of the principal diagonals of the matrix $\mathbf C$ gives the relative standard deviation for each fitted parameter. Of course, all the relevant sensitivity functions must be defined for the parameter set $P$ obtained at the end of the iteration.

\subsection{Thermal behaviour of the reinforced metal foams}

The flash experiments were performed multiple times on the presented metal foam samples, and the measured data were evaluated by the iterative procedure described, using both the Fourier and the GK heat equations. The resulting parameters are listed in Table~\ref{tab:metalfoams} for each sample. While under the representative sample size, besides size dependence, statistical variation between the samples can be present as well and influence the parameters. This is visible in Table~\ref{tab:metalfoams}, e.g., for samples 22 and 24 as the same types of specimen exhibited different values of certain thermal parameters.

\begin{figure}
	\centering
	\includegraphics[width=7.65cm]{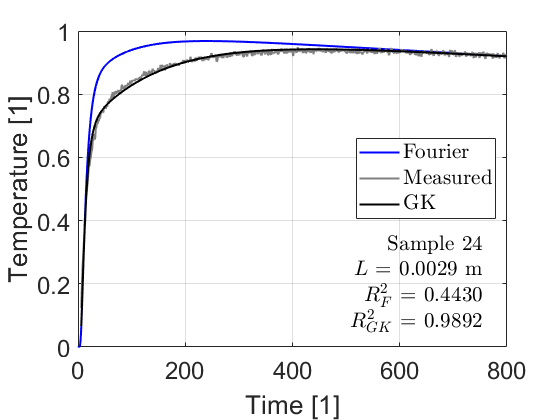}
	\includegraphics[width=7.65cm]{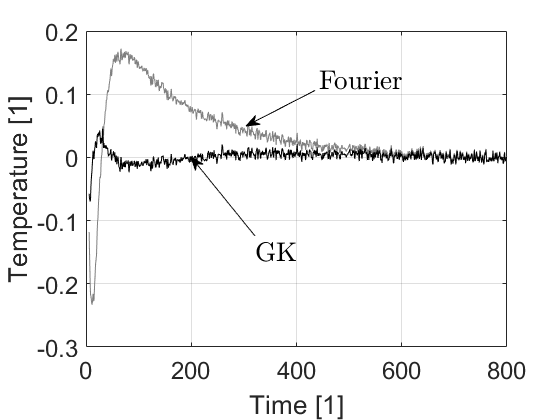}
	\caption{Evaluation of metal foam (sample 24) and difference between the evaluation curve and the measured temperature.}
	\label{fig:femhab-plot}
\end{figure}

\begin{figure}
	\centering
	\includegraphics[width=7.65cm]{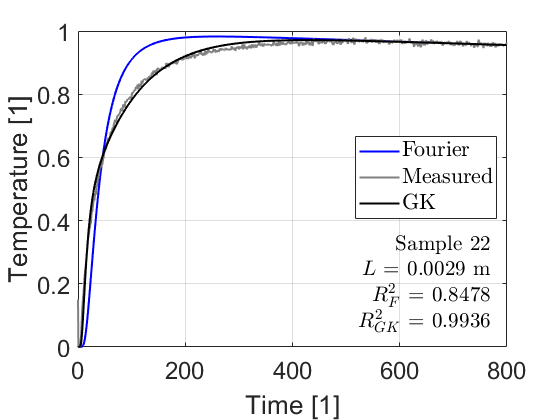}
	\includegraphics[width=7.65cm]{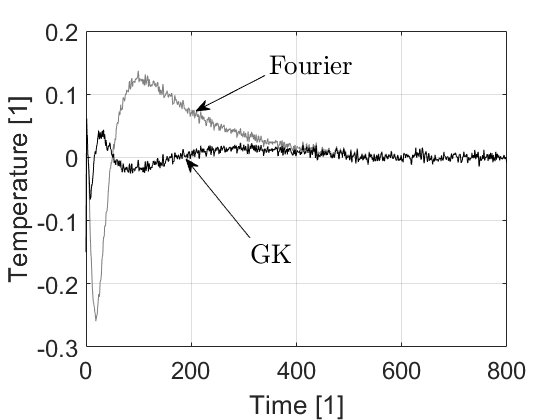}
	\caption{Evaluation of metal foam (sample 22) and difference between the evaluation curve and the measured temperature.}
	\label{fig:femhab22-plot}
\end{figure}

All the samples, regardless of the type and amount of reinforcing material, have shown a non-Fourier effect, i.e., the Fourier heat equation cannot properly model the measured temperature history. Shortly after the heat pulse, the measured temperature rise is faster than Fourier's prediction. After that, the characteristics exchange, Fourier's prediction significantly overshoots the measured one. This is a clear appearance of multiple time scales, and called over-diffusion. 
On contrary, when testing the GK heat equation, it can be clearly seen from Figs.~\ref{fig:femhab-plot} and~\ref{fig:femhab22-plot} that a two time scale model can effectively reproduce the measurements, providing a notably better fit. In accordance, the temperature difference curves ($T_{\textrm{diff}}$) are also insightful (right side of Figs.~\ref{fig:femhab-plot} and~\ref{fig:femhab22-plot}), though the use of the GK equation can considerably reduce the errors, the appearance of a third time scale is also visible. This is reasonable as these specimens have a complex structure, let us recall that the ceramic sphere are hollow, filled with air, therefore heat conduction, convection, and radiation are all influencing the outcome. 

\begin{figure}[H]
	\centering
	\includegraphics[width=17cm]{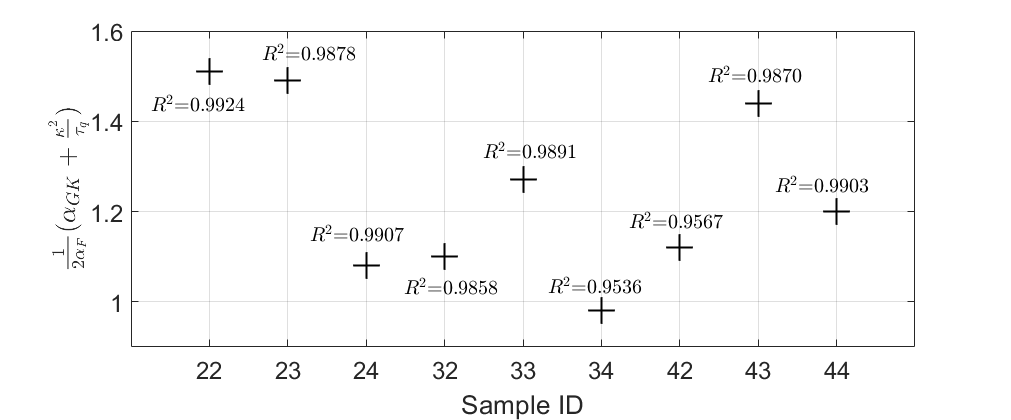}
	\caption{Deviation of samples from Fourier resonance.}
	\label{fig:atlag}
\end{figure}

We want to call the attention to Fig.~\ref{fig:atlag}. It shows how the fitted parameters of both heat equations relate to each other. This comparison underlines the importance of Eq.~\eqref{eq:atlag}, viz., how to determine reliably the Fourier thermal diffusivity. While Fourier's equation fits well the slow cooling interval of the measured data, the essential transient part cannot be evaluated reliably, therefore the corresponding thermal diffusivity $\alpha_F$ is also seriously uncertain in such a case. That is, despite the seeming coincidence between the Fourier and GK heat equations in the cooling interval, it does not ensure that the corresponding thermal diffusivity found with Fourier's law can be utilized even with limitations since the conduction part is practically zero in such a slow process (let us recall that this is the reason why we can use only one heat transfer boundary condition), hence Fourier's heat equation can describe the cooling domain with almost any thermal diffusivity, principally. However, what we propose here is to utilise Eq.~\eqref{eq:atlag} to determine the corresponding Fourier thermal diffusivity, represented under the 'Calculated $\alpha_{Fc}$' in Table \ref{tab:metalfoams}.

\begin{table}[]
	\begin{tabular}{ c c c ccc ccc c }
		\toprule
		&                     & \multicolumn{1}{c}{Fourier}                         & \multicolumn{3}{c}{Guyer--Krumhansl}                                                                                              & \multicolumn{3}{c}{Error values}                                                                                                           &                                                                                   \\ \cmidrule{3-9}
		& \multirow{-2}{*}{L} & \multicolumn{1}{c}{$\alpha_{Fc}$}                      & \multicolumn{1}{c}{$\alpha_{GK}$}                   & \multicolumn{1}{c}{$\tau_q$} & \multicolumn{1}{c}{$\kappa^2$}            & \multicolumn{1}{c}{$D$}                           & \multicolumn{1}{c}{$\alpha$}                        & \multicolumn{1}{c}{$\tau_q$} & \multirow{-2}{*}{\begin{tabular}[c]{@{}c@{}}Calculated\\ $\alpha_F$\end{tabular}} \\ \cmidrule{2-10} 
		\multirow{-3}{*}{Sample ID} & {[}m{]}             & \multicolumn{1}{c}{$10^{-6}$ {[}$\frac{m^2}{s}${]}} & \multicolumn{1}{c}{$10^{-6}$ {[}$\frac{m^2}{s}${]}} & \multicolumn{1}{c}{{[}s{]}}  & \multicolumn{1}{c}{$10^{-6}$ {[}$m^2${]}} & \multicolumn{1}{c}{$10^{-8}$ {[}$\frac{m^2}{s}${]}} & \multicolumn{1}{c}{$10^{-9}$ {[}$\frac{m^2}{s}${]}} & \multicolumn{1}{c}{{[}s{]}}  & $10^{-6}$ {[}$\frac{m^2}{s}${]}                                                   \\ \midrule
		22                          & 0.0029              & 2.59                                                 & \multicolumn{1}{c}{1.92}                            & \multicolumn{1}{c}{0.21}     & 1.28                                       & \multicolumn{1}{c}{\cellcolor[HTML]{FFFFFF}5.15}    & \multicolumn{1}{c}{6.46}                            & 0.003                         & 3.92                                                                              \\ \midrule
		23                          & 0.0020               & 1.19                                                 & \multicolumn{1}{c}{0.88}                            & \multicolumn{1}{c}{0.23}     & 0.62                                       & \multicolumn{1}{c}{\cellcolor[HTML]{FFFFFF}1.94}    & \multicolumn{1}{c}{2.56}                            & 0.003                         & 1.77                                                                              \\ \midrule
		24                          & 0.0029              & 5.5                                                  & \multicolumn{1}{c}{2.87}                            & \multicolumn{1}{c}{0.29}     & 2.64                                       & \multicolumn{1}{c}{\cellcolor[HTML]{FFFFFF}4.30}    & \multicolumn{1}{c}{9.78}                            & 0.003                         & 5.96                                                                              \\ \midrule
		32                          & 0.0021              & 2.85                                                 & \multicolumn{1}{c}{1.56}                            & \multicolumn{1}{c}{0.28}     & 1.31                                       & \multicolumn{1}{c}{\cellcolor[HTML]{FFFFFF}2.65}    & \multicolumn{1}{c}{6.18}                            & 0.003                         & 3.15                                                                              \\ \midrule
		33                          & 0.0021              & 1.73                                                 & \multicolumn{1}{c}{1.07}                            & \multicolumn{1}{c}{0.30}     & 1.01                                       & \multicolumn{1}{c}{\cellcolor[HTML]{FFFFFF}1.92}    & \multicolumn{1}{c}{3.44}                            & 0.003                         & 2.20                                                                               \\ \midrule
		34                          & 0.0020               & 3.47                                                 & \multicolumn{1}{c}{2.04}                            & \multicolumn{1}{c}{0.26}     & 1.52                                       & \multicolumn{1}{c}{\cellcolor[HTML]{FFFFFF}2.47}    & \multicolumn{1}{c}{8.48}                            & 0.004                         & 2.83                                                                              \\ \midrule
		42                          & 0.0028              & 5.41                                                 & \multicolumn{1}{c}{4.25}                            & \multicolumn{1}{c}{0.22}     & 1.62                                       & \multicolumn{1}{c}{\cellcolor[HTML]{FFFFFF}9.19}    & \multicolumn{1}{c}{23.1}                            & 0.005                         & 6.08                                                                              \\ \midrule
		43                          & 0.0020               & 2.23                                                 & \multicolumn{1}{c}{1.26}                            & \multicolumn{1}{c}{0.18}     & 0.91                                       & \multicolumn{1}{c}{\cellcolor[HTML]{FFFFFF}3.77}    & \multicolumn{1}{c}{4.43}                            & 0.002                         & 3.20                                                                               \\ \midrule
		44                          & 0.0035              & 4.57                                                 & \multicolumn{1}{c}{2.65}                            & \multicolumn{1}{c}{0.37}     & 3.07                                       & \multicolumn{1}{c}{\cellcolor[HTML]{FFFFFF}5.05}    & \multicolumn{1}{c}{9.89}                            & 0.004                         & 5.48                                                                              \\ \bottomrule
	\end{tabular}
\caption{Summary of iteratively fitted thermal parameters of the metal foams.}\label{tab:metalfoams}
\end{table}

With knowledge of the composition of the metal foams, the effective thermal diffusivity can be calculated for each sample. After calculation, values of an order of magnitude higher for each sample type were obtained. This difference shows very clearly how significantly the thermal properties are reduced, due to the thermal contact resistances around the reinforcing materials, while also depending on the cavity size of the reinforcing material. 

\begin{figure}[H]
	\centering
	\includegraphics[width=7.65cm]{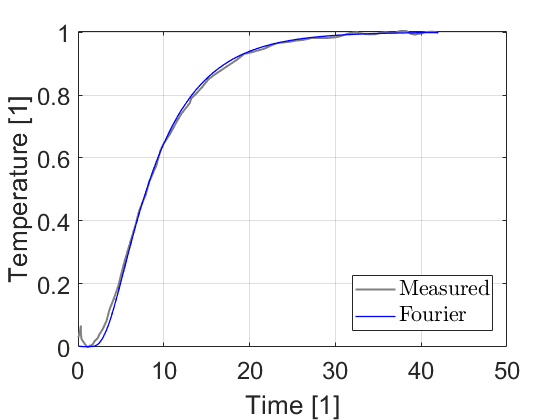}
    \caption{Evaluation of a metal foam sample in \cite{lunev_digital_2022} using an iterative procedure.}
	\label{fig:artem}
\end{figure}

Briefly, we would like to have a remark on the measurements presented in Ref.~\cite{lunev_digital_2022}, performed by A.~Lunev et al. Their measurements are also performed on metal foams with various porosity, using a standardized commercial laser flash apparatus (LFA). They also encountered similar deviations from Fourier's law, although the differences are not as strong as here. They also concluded that Fourier's law is not able to provide such an effective parameter, which describes the overall transient behaviour of such a complex material structure in such a size and under such transients. Its importance is that commercial LFA equipment is used to determine the thermal diffusivity as a standard technique in engineering, but a representative sample size would be much larger than the current limits of usual LFA devices.
For curiosity, we also tested the previously described iterative evaluation procedure on the  data they published, and confirmed to be different from Fourier's. However, we found that the deviation between Fourier's and the measured curves are so small that the iterative GK evaluation method quits the iteration immediately as the resulting parameters are close to the Fourier resonance condition $\alpha = \kappa^2/\tau_q$, therefore the corresponding sensitivities are linearly dependent, and the determinant of the matrix $\hat{\mathbf S}^T \hat{\mathbf S}$ becomes zero. Using Fourier's law, we obtained the same fit as in \cite{lunev_digital_2022}, see Fig.~\ref{fig:artem} for details.

\subsection{Effective thermal conductivity}

For a sample with known inner structure, therefore with known volume fractions for each component, it is possible to calculate the effective density and specific heat since these quantities do not depend on the interface properties such as thermal resistances. However, the effective thermal conductivity is much more challenging to determine, and therefore, one can solely estimate its value. There are numerous estimations notably exploiting the knowledge about the material structure \cite{FiedlerEtal14}. All these estimation procedures and the related (semi-)empirical formulae are strongly limited. 

Here, we perform a brief but insightful comparison of effective thermal conductivity estimations. Table \ref{tab:effectiv} presents the calculated values of mass density and specific heat. Additionally, Table \ref{tab:effectiv} consists of the corresponding thermal diffusivity values found using the Fourier and GK heat equations, where  $\alpha_{Fc}$ is obtained from the averaging the GK coefficients based on~\eqref{eq:atlag}, and the subscript "c" aims to emphasise it. Consequently, $\lambda_{Fc}$ follows from $\alpha_{Fc}$.

\begin{table}[H]
	\begin{tabular}{ccccccc}
		\hline
		\multirow{2}{*}{Sample ID} & $\rho$                 & $c$                   & $\alpha_{Fc}$                                      & $\alpha_{GK}$                                     & $\lambda_{Fc}$          & $\lambda_{GK}$         \\ \cline{2-7} 
		& {[}$\frac{kg}{m^3}${]} & {[}$\frac{J}{kgK}${]} & $10^{-6}{[}\frac{m^2}{s}${]} & $10^{-6}{[}\frac{m^2}{s}${]} & {[}$\frac{W}{mK}${]} & {[}$\frac{W}{mK}${]} \\ \hline
		2x                         & 1650                   & 767.5                 & 6.17                                            & 4.88                                           & 7.82                 & 6.17                 \\ \hline
		3x                         & 1738                   & 768.2                 & 2.99                                           & 1.52                                            & 3.99                 & 2.02                 \\ \hline
		4x                         & 1738                   & 768.2                 & 4.78                                            & 2.41                                            & 6.38                 & 3.21                \\ \hline
	\end{tabular}
\caption{Summary of sample properties.}\label{tab:effectiv}
\end{table}

The effective thermal conductivity can be estimated in other ways, three of which are mentioned below, only for demonstrative purposes. We use the following estimations:

\begin{align}
&\textrm{Voigt-type:}&	\quad \lambda_{eff} = V_1 \lambda_1 + V_2 \lambda_2; \\
&\textrm{Markworth et al.:}& \quad \lambda_{eff} = V_1 \lambda_1 + V_2 \lambda_2 + V_1 V_2 \frac{\lambda_1-\lambda_2}{\frac{3}{\frac{\lambda_2}{\lambda_1}-1}+V_1}; \\
&\textrm{Wakashima--Tsukamoto:}& \quad \lambda_{eff} = \lambda_1 + \frac{\lambda_1 V_2 (\lambda_2-\lambda_1)}{\lambda_1 + \frac{(\lambda_1-\lambda_2) V_1}{3}},
\end{align}
where $V_1$ and $V_2 = 1 - V_1$ are the corresponding volume fractions of the components, and $\lambda_1$ and $\lambda_2$ are the thermal conductivity of the corresponding component. The effective thermal conductivity calculated for each foam type is given in table \ref{tab:effectiv_calc} for each approach. As the studies foams consist of four components, these were reduced to 1-1 pair for the estimations with the same formula. 

These calculations show that  each approach significantly over-estimates the actual measured value. This is reasonable since both the CHSs and the reinforcing particles introduce considerable amount of contact thermal resistance; their overall volume fraction is not representative enough.

\begin{table}[H]
	\begin{tabular}{cccc}
		\hline
		\multirow{2}{*}{Sample ID} & \multicolumn{3}{c}{$\lambda_{eff}$ {[}$\frac{W}{mK}${]}}  \\ \cline{2-4} 
		& Voigt-type  & Markworth & Wakashima--Tsukamoto \\ \hline
		2x                         & 54.52             & 39.79     & 66.17               \\ \hline
		3x and 4x                  & 55.82             & 41.44     & 67.19               \\ \hline
	\end{tabular}
\caption{Summary of calculated effective thermal conductivity.}\label{tab:effectiv_calc}
\end{table}

\section{Summary}

The thermal behaviour of AlSi7Mg metal matrix syntactic foams with ceramic hollow sphere phase fillers, and \ch{SiC} or \ch{Al2O3} reinforcing particles has been investigated. Heat pulse experiments were performed in order to characterize the thermal diffusive behaviour of the samples, and calculate their thermal parameters. The material's inherent heterogeneous structure necessitated the use of a heat conduction model beyond Fourier's law, in the form of the Guyer--Krumhansl equation. The evaluation of the measurement data using the GK model has shown a significantly better improvement over Fourier's equation, while the Fourier model is unable to reliably provide the necessary thermal diffusivity based on the transient temperature history. The iterative parameter fitting method developed for the GK equation is proved be an efficient approach to determine the model parameters, and also to insert the model attributes through the sensitivity functions.

The measured thermal diffusivity $\alpha$ (Fourier and GK), heat flux relaxation time $\tau_q$, and spatial scale parameter $\kappa^2$ (GK) values have been given for each sample type and size of applied phase filler materials. A shortcoming of the measurement method is the small size of the specimens that can be tested, which resulted in a discrepancy in the measured parameters between similar samples. Furthermore, due to the statistical variations in the reinforcing particles, even the same type of samples can result in different parameter values (although the parameters fall in the same order of magnitude), and that requires to investigate a considerably larger number of samples. 
The consistent variation of composition would also be insightful in regard the role of relaxation time and spatial scale parameter to understand their currently undiscovered meaning. That would enable the design of specific heat conduction parameters.
In conclusion, the GK model clearly allows for an accurate description of the thermal behaviour of the material investigated using bulk parameters without the need for microscopic, pore-scale simulations.

For an even more detailed description, measuring the density and specific heat capacity of the material could also be of interest. This requires a separate experimental investigation, which might be explored in the future. Additionally, improvements to the heat pulse experiments which would allow for larger samples would be worthwhile to explore.

\section*{Acknowledgements}
Project no.~TKP-6-6/PALY-2021 has been implemented with the support provided by the Ministry of Culture and Innovation of Hungary from the National Research, Development and Innovation Fund, financed under the TKP2021-NVA funding scheme (A.F.). The research was funded by the Sustainable Development and Technologies National Programme of the Hungarian Academy of Sciences (FFT NP FTA) (R.K.). This work was partially supported in part by the Hungarian Scientific Research Fund under Grant agreements OTKA 138505 (I.O., J.M.) and 134277 (R.K.). This work was performed in the frame of the 2021-2.1.2-HŐ-2021-00004 project, implemented with the support provided by the National Research, Development and Innovation Fund of Hungary, financed under the 2021-2.1.2-HŐ funding scheme (D.T.).

\section*{Author contributions}
A. Fehér: thermal measurements, development of the iteration procedure, manuscript preparation. J. E. Maróti: investigation, validation. D. M. Takács: writing and editing. I. N. Orbulov and R. Kovács: resources, validation, writing –  review \& editing, supervision.

\bibliographystyle{unsrt}

\begin{thebibliography}{100}
	
	\bibitem{banhart_aluminium_2005}
	John Banhart.
	\newblock Aluminium foams for lighter vehicles.
	\newblock {\em International Journal of Vehicle Design}, 37(2/3):114, 2005.
	
	\bibitem{song_experimental_2020}
	Jiafeng Song, Shucai Xu, Lihan Xu, Jianfei Zhou, and Meng Zou.
	\newblock Experimental study on the crashworthiness of bio-inspired aluminum foam-filled tubes under axial compression loading.
	\newblock {\em Thin-Walled Structures}, 155:106937, October 2020.
	
	\bibitem{srinath_characteristics_2010}
	G.~Srinath, Aravind Vadiraj, G.~Balachandran, S.~N. Sahu, and Amol~A. Gokhale.
	\newblock Characteristics of aluminium metal foam for automotive applications.
	\newblock {\em Transactions of the Indian Institute of Metals}, 63(5):765--772, October 2010.
	
	\bibitem{linul_axial_2021}
	Emanoil Linul and Omid Khezrzadeh.
	\newblock Axial crashworthiness performance of foam-based composite structures under extreme temperature conditions.
	\newblock {\em Composite Structures}, 271:114156, September 2021.
	
	\bibitem{claar_ultra-lightweight_2000}
	T.~Dennis Claar, Chin-Jye Yu, Ian Hall, John Banhart, Joachim Baumeister, and Wolfgang Seeliger.
	\newblock Ultra-{Lightweight} {Aluminum} {Foam} {Materials} for {Automotive} {Applications}.
	\newblock pages 2000--01--0335, March 2000.
	
	\bibitem{simancik_metallic_2001}
	F.~Simancik.
	\newblock Metallic foams - ultra light materials for structural applications.
	\newblock {\em Inżynieria Materiałowa}, (R. XXII, nr 5):823--828, 2001.
	
	\bibitem{lefebvre_porous_2008}
	L.-P. Lefebvre, J.~Banhart, and D.~C. Dunand.
	\newblock Porous {Metals} and {Metallic} {Foams}: {Current} {Status} and {Recent} {Developments}.
	\newblock {\em Advanced Engineering Materials}, 10(9):775--787, September 2008.
	
	\bibitem{jones_assessment_2009}
	Michael Jones, Tony Parrott, Daniel Sutliff, and Christopher Hughes.
	\newblock Assessment of {Soft} {Vane} and {Metal} {Foam} {Engine} {Noise} {Reduction} {Concepts}.
	\newblock In {\em 15th {AIAA}/{CEAS} {Aeroacoustics} {Conference} (30th {AIAA} {Aeroacoustics} {Conference})}, Miami, Florida, May 2009. American Institute of Aeronautics and Astronautics.
	
	\bibitem{liu_prediction_2012}
	Hanru Liu, Jinjia Wei, and Zhiguo Qu.
	\newblock Prediction of aerodynamic noise reduction by using open-cell metal foam.
	\newblock {\em Journal of Sound and Vibration}, 331(7):1483--1497, March 2012.
	
	\bibitem{klavzar_protective_2015}
	Andreas Klavzar, Maxime Chiroli, Anne Jung, and Bernhard Reck.
	\newblock Protective {Performance} of {Hybrid} {Metal} {Foams} as {MMOD} {Shields}.
	\newblock {\em Procedia Engineering}, 103:294--301, 2015.
	
	\bibitem{williamsen_video_2001}
	Joel Williamsen and Eric Howard.
	\newblock Video imaging of debris clouds following penetration of lightweight spacecraft materials.
	\newblock {\em International Journal of Impact Engineering}, 26(1-10):865--877, December 2001.
	
	\bibitem{ryan_hypervelocity_2010}
	Shannon Ryan, E~Ordonez, EL~Christiansen, and DM~Lear.
	\newblock Hypervelocity impact performance of open cell foam core sandwich panel structures.
	\newblock In {\em Hypervelocity {Impact} {Symposium}}, 2010.
	\newblock Issue: JSC-CN-19432.
	
	\bibitem{rajak_insight_2020}
	Dipen~Kumar Rajak and Manoj Gupta.
	\newblock {\em An {Insight} {Into} {Metal} {Based} {Foams}: {Processing}, {Properties} and {Applications}}, volume 145 of {\em Advanced {Structured} {Materials}}.
	\newblock Springer Singapore, Singapore, 2020.
	
	\bibitem{gupta_metal_2014}
	Nikhil Gupta and Pradeep~K Rohatgi.
	\newblock {\em Metal matrix syntactic foams: processing, microstructure, properties and applications}.
	\newblock DEStech Publications, Inc, 2014.
	
	\bibitem{cochran_ceramic_1998}
	Joe~K Cochran.
	\newblock Ceramic hollow spheres and their applications.
	\newblock {\em Current Opinion in Solid State and Materials Science}, 3(5):474--479, October 1998.
	
	\bibitem{wright_processing_2017}
	Andrew Wright and Andrew Kennedy.
	\newblock The {Processing} and {Properties} of {Syntactic} {Al} {Foams} {Containing} {Low} {Cost} {Expanded} {Glass} {Particles}: {Processing} and {Properties} of {Syntactic} {Al} {Foams}.
	\newblock {\em Advanced Engineering Materials}, 19(11):1600467, November 2017.
	
	\bibitem{szlancsik_mechanical_2020}
	A~Szlancsik, D~Kincses, and I~N Orbulov.
	\newblock Mechanical properties of {AlSi10MnMg} matrix syntactic foams filled with lightweight expanded clay particles.
	\newblock {\em IOP Conference Series: Materials Science and Engineering}, 903(1):012045, August 2020.
	
	\bibitem{stevenson_particle_2012}
	G.~Kaptay and N.~Babcsán.
	\newblock Particle {Stabilized} {Foams}.
	\newblock In Paul Stevenson, editor, {\em Foam {Engineering}}, pages 121--143. Wiley, 1 edition, February 2012.
	
	\bibitem{maroti_characteristic_2023}
	János~Endre Maróti and Imre~Norbert Orbulov.
	\newblock Characteristic compressive properties of {AlSi7Mg} matrix syntactic foams reinforced by {Al2O3} or {SiC} particles in the matrix.
	\newblock {\em Materials Science and Engineering: A}, 869:144817, March 2023.
	
	\bibitem{boomsma_metal_2003}
	K.~Boomsma, D.~Poulikakos, and F.~Zwick.
	\newblock Metal foams as compact high performance heat exchangers.
	\newblock {\em Mechanics of Materials}, 35(12):1161--1176, December 2003.
	
	\bibitem{ejlali_application_2009}
	Azadeh Ejlali, Arash Ejlali, Kamel Hooman, and Hal Gurgenci.
	\newblock Application of high porosity metal foams as air-cooled heat exchangers to high heat load removal systems.
	\newblock {\em International Communications in Heat and Mass Transfer}, 36(7):674--679, August 2009.
	
	\bibitem{jannesari_experimental_2017}
	Hamid Jannesari and Naeim Abdollahi.
	\newblock Experimental and numerical study of thin ring and annular fin effects on improving the ice formation in ice-on-coil thermal storage systems.
	\newblock {\em Applied Energy}, 189:369--384, March 2017.
	
	\bibitem{chen_thermal_2021}
	Xue Chen, Xiaolei Li, Xinlin Xia, Chuang Sun, and Rongqiang Liu.
	\newblock Thermal storage analysis of a foam-filled {PCM} heat exchanger subjected to fluctuating flow conditions.
	\newblock {\em Energy}, 216:119259, February 2021.
	
	\bibitem{talebizadehsardari_consecutive_2021}
	Pouyan Talebizadehsardari, Jasim~M. Mahdi, Hayder~I. Mohammed, M.A. Moghimi, Amir Hossein~Eisapour, and Mohammad Ghalambaz.
	\newblock Consecutive charging and discharging of a {PCM}-based plate heat exchanger with zigzag configuration.
	\newblock {\em Applied Thermal Engineering}, 193:116970, July 2021.
	
	\bibitem{shu_effect_2023}
	Gao Shu, Tian Xiao, Junfei Guo, Pan Wei, Xiaohu Yang, and Ya-Ling He.
	\newblock Effect of charging/discharging temperatures upon melting and solidification of {PCM}-metal foam composite in a heat storage tube.
	\newblock {\em International Journal of Heat and Mass Transfer}, 201:123555, February 2023.
	
	\bibitem{tian_numerical_2011}
	Y.~Tian and C.Y. Zhao.
	\newblock A numerical investigation of heat transfer in phase change materials ({PCMs}) embedded in porous metals.
	\newblock {\em Energy}, 36(9):5539--5546, September 2011.
	
	\bibitem{nemati_pore-scale_2023}
	Hossain Nemati, Vahid Souriaee, Meysam Habibi, and Kambiz Vafai.
	\newblock Pore-scale and volume-averaged simulations of phase change material melting: {A} comparison between local and nonlocal thermal equilibrium.
	\newblock {\em Numerical Heat Transfer, Part A: Applications}, pages 1--15, March 2023.
	
	\bibitem{both_deviation_2016}
	Soma Both, Balázs Czél, Tamás Fülöp, Gyula Gróf, Ákos Gyenis, Róbert Kovács, Peter Ván, and József Verhás.
	\newblock Deviation from the {Fourier} law in room-temperature heat pulse experiments.
	\newblock {\em Journal of Non-Equilibrium Thermodynamics}, 41(1):41--48, January 2016.
	
	\bibitem{van_guyer-krumhansltype_2017}
	P.~Ván, A.~Berezovski, T.~Fülöp, Gy. Gróf, R.~Kovács, Á. Lovas, and J.~Verhás.
	\newblock Guyer-{Krumhansl}–type heat conduction at room temperature.
	\newblock {\em EPL (Europhysics Letters)}, 118(5):50005, June 2017.
	
	\bibitem{fulop_emergence_2018}
	Tamás Fülöp, Róbert Kovács, Ádám Lovas, Ágnes Rieth, Tamás Fodor, Mátyás Szücs, Péter Ván, and Gyula Gróf.
	\newblock Emergence of {Non}-{Fourier} {Hierarchies}.
	\newblock {\em Entropy}, 20(11):832, October 2018.
	
	\bibitem{feher_evaluation_2021}
	A.~Fehér and R.~Kovács.
	\newblock On the evaluation of non-{Fourier} effects in heat pulse experiments.
	\newblock {\em International Journal of Engineering Science}, 169:103577, December 2021.
	
	\bibitem{lunev_digital_2022}
	Artem Lunev, Alexander Lauerer, Vadim Zborovskii, and Fabien Léonard.
	\newblock Digital twin of a laser flash experiment helps to assess the thermal performance of metal foams.
	\newblock {\em International Journal of Thermal Sciences}, 181:107743, November 2022.
	
	\bibitem{guyer_thermal_1966}
	R.~A. Guyer and J.~A. Krumhansl.
	\newblock Thermal {Conductivity}, {Second} {Sound}, and {Phonon} {Hydrodynamic} {Phenomena} in {Nonmetallic} {Crystals}.
	\newblock {\em Physical Review}, 148(2):778--788, August 1966.
	
	\bibitem{feher_analytical_2022}
	Anna Fehér and Róbert Kovács.
	\newblock Analytical evaluation of non-{Fourier} heat pulse experiments on room temperature.
	\newblock {\em IFAC-PapersOnLine}, 55(18):87--92, 2022.
	
	\bibitem{fabrizio_stability_2014}
	Mauro Fabrizio and Barbara Lazzari.
	\newblock Stability and {Second} {Law} of {Thermodynamics} in dual-phase-lag heat conduction.
	\newblock {\em International Journal of Heat and Mass Transfer}, 74:484--489, July 2014.
	
	\bibitem{isotc_164sc_2_ductility_testing_technical_committee_mechanical_2011}
	{ISO/TC 164/SC 2 Ductility testing Technical Committee}.
	\newblock Mechanical testing of metals – {Ductility} testing – {Compression} test for porous and cellular metals.
	\newblock Standard ISO 13314:2011, International Organization for Standardization, Geneva, CH, 2011.
	
	\bibitem{le_masson_metti_2015}
	Pascal Le~Masson, Olivier Fudym, Jean-Laurent Gardarein, and Denis Maillet.
	\newblock Metti 6 {Advanced} {School}: {Thermal} {Measurements} and {Inverse} {Techniques}.
	\newblock Lecture notes, 2015.
	
	\bibitem{FiedlerEtal14}
	T.~Fiedler, M.~A.~Sulong, M.~Vesenjak, Y.~Higa, I.~V.~Belova, A.~{\"O}chsner, and G.~E.~Murch.
	\newblock Determination of the thermal conductivity of periodic APM foam models.
	\newblock {\em International Journal of Heat and Mass Transfer}, 73:826--833, 2014.
	
\end{thebibliography}

\end{document}